\documentstyle[aps,preprint,tighten]{revtex}

\input{epsf}
\def\Journal#1#2#3#4{{#1} {\bf #2}, #3 (#4)}

\def\NPB{{\em Nucl. Phys.} B}
\def\PLB{{\em Phys. Lett.}  B}
\def\PRL{\em Phys. Rev. Lett.}
\def\PRD{{\em Phys. Rev.} D}

\begin{document}

\title{The Spectral Density of the Dirac Operator above $T_c$Rep}

\author{Thomas D. Cohen}

\address{Department of 
Physics, University of~Maryland, College~Park, MD~20742}

\maketitle

\begin{abstract}

The importance of the spectral density of the Dirac operator in studying 
spontaneous chiral symmetry breaking  and anomalous $U(1)_A$ symmetry breaking
are reviewed.  It is shown that both types of symmetry breaking can be traced
to effects of modes near zero virtuality.  Above $T_c$, where chiral symmetry is
restored, it is shown on general grounds that (in the $m_q \rightarrow 0$
limit), the density of states vanishes at zero
virtuality faster than $\lambda$, where $\lambda$ is the virtuality---{\it
i.e.} $\rho(\lambda) \sim |\lambda|^\alpha$ is not possible
for $\alpha \le 1$. Isospin invariance is used to show that $\rho(\lambda)
\sim m_q^{1-\alpha} |\lambda|^\alpha$ is
also not possible for $\alpha \le 1$.  State-of-the-art lattice calculations are reviewed in light
of  these constraints.  In particular, it is  argued that violations of
these constraints by lattice calculations indicate possible large systematic 
errors; this  raises questions about $U(1)_A$  violating  effects seen on the lattice.
It is also shown that above $T_c$, the Dirac spectrum 
has a gap near zero (in the $m_q \rightarrow 0$ limit) unless contributions
from quark-line--connected and disconnected contributions conspire to cancel.
\vspace{.5in} \\  \\
 \end{abstract} 

\section{Introduction}
It is a great pleasure to give this talk at a workshop in honor of Mannque
Rho's $60^{\rm th}$ birthday. Mannque has been a central figure in our field for 
quite a long time.   When I began working in nuclear physics about 15 years 
ago, it became  quite clear to me  that Mannque was quite special.
I soon realized that  our field could be roughly divided into two
classes---theoretical physicists with powerful and elegant
mathematical tools at their disposal, and those with a phenomenological bent
who had the intuition to make the kinds of simplifications needed to deal 
with the
complications of realistic strongly interacting systems and make experimentally
relevent predictions.  Mannque is almost unique
in belonging to  both groups.

Having divided the field this way,  let me begin this talk with an apology. 
Virtually nothing which I will deal with here has immediate phenomenological consequences.
The problem with which I am dealing concerns the QCD vacuum 
in thermal equilibrium above the chiral phase transition.  This raises the
obvious question of just how does one heat up vacuum?
 An idealized experimental setup---as designed by
a typical American student---consists of an empty test tube held above a large candle.  I have been told that this method will not work in practice,
and that the only practical way to explore this phase in the laboratory 
is through ultrarelativistic heavy ion collisions.
Unfortunately, the way the physics of the phase of interest
gets translated into experimental signatures which  can be observed in heavy 
ion reactions is by no means clear.  I will barely touch  on
this question in this talk, and what I do suggest will turn out to be highly
sensitive to assumptions about the dynamics of the phase transition and will be
{\it highly} speculative.

A theorist might wish to attack a problem so far removed from
experiment for reasons other than the sheer delight in being unable to be shown
wrong by our grungy experimental friends.
The fact is that we believe we 
know the underlying theory (QCD) and we can cleanly
formulate well-posed questions about QCD in thermal equilibrium in terms of the
degrees of freedom of QCD.  Moreover, at least in principle, these
questions can be answered to arbitrary precision via sufficiently good lattice
simulations.  Of course in practice, present day lattice simulations 
have some serious shortcomings for technical reasons associated with a lack of
sufficient computer power, but there is every reason to believe that as
computers get better,  the lattice community will slowly converge on the right answers.

Since we are ultimately concerned with issues of nonperturbative QCD, but we
will be working with the QCD degrees of freedom---quarks  and gluon--it is probably useful to
introduce a few tools.  One essential tool is the composite operator
or interpolating field.  These beasts are constructed from the quark and gluon
operators.  Typically one is interested in local gauge-invariant operators with
fixed transformation properties under Lorentz transformations, parity, and
isospin rotations.  I will generically note such operators as $J$ and examples
include:
\begin{eqnarray}
J & \sim & \overline{q} \, i  \gamma_5 \tau \,q \nonumber\\
J & \sim & F_{\mu \nu} F^{\mu \nu}
\end{eqnarray}
which have the quantum numbers of a pseudoscalar isovector and scalar isoscalar,
respectively.  One can construct an infinite number of such operators.
Essentially all of the information one can extract from QCD is encoded in the
various n-point correlation functions of these currents.  A typical example
is the two-point function $\langle J(x,t) J(0,0) \rangle$ where one can choose
the appropriate boundary conditions of interest ({\it e.g.}, time-ordered,
advanced, retarded, {\it etc.})  These correlators tell us about the response of the QCD vacuum
given a disturbance associated with $J$.  

How can one calculate these things?  In principle, there is a straightforward way
to calculate the expectation value of any observable given in terms of quark
and gluon field operators.  The method is functional integration.  Here we
will discuss the problem for imaginary time (Euclidean space) and ``all'' one
needs to do to get real time results is analytically continue.  The Euclidean
functional integral for the expectation value of any operator  is given by
\begin{eqnarray}
\langle {\cal O} \rangle \, & = &\, \frac{1}{Z} \int \, [D A]  \, 
[D \overline{q}] \, [D q] \, e^{i S_{\rm QCD}}\, {\cal O} \nonumber \\
Z &= & \int \, [D A] \, [D \overline{q}] \, [D q] \, e^{i S_{\rm QCD}}
\label{func1} \end{eqnarray}
where $S_{QCD}$ is the action.

Here we are interested in QCD at finite temperatures.  One of the virtues
of the Euclidean space functional integral is that one can calculate finite
thermal expectation values exactly as one calculates zero temperature ones. 
Indeed the functional integral expression in eqs.~(\ref{func1}) holds 
for thermal expectation values.  The only difference from the free-space case
concerns the class of configurations over which one integrates.  To use
functional integrals at finite
temperatures one simply imposes the boundary condition\cite{ft} that all boson fields
(gluons in this case)  are periodic in time with a period  given by $\beta$,
the inverse temperature, while fermion field configurations are anti-periodic:
$$A_{\mu}(x,t) \, = \, A_{\mu}(x, t+ \beta) \; \; \; \;  
q(x,t) \, = \,-q(x, t+ \beta)$$

The above formulation holds for any operator.  The operator which I will
 stress in this talk is the density of states of the
Dirac operator; the question I will address is how this relates to
various correlation functions.  The operator can be defined in the following
way:
\begin{equation}
\rho (\lambda)_A \, = 1/{\rm Vol} \, \, \sum_{j} \delta (\lambda - \lambda_j)
\end{equation}
where $i \lambda_j$ is  an eigenvector of the (Euclidean) covariant Dirac
operator acting on spinors,
$$D_\mu \gamma^\mu q_j \, = \,i \lambda_j q_j\,,$$
and Vol is the volume of space-time with a finite temporal extent given by
$\beta$ and, ultimately, infinite spatial extent.
Note that the  Dirac operator $D_\mu \gamma^\mu$ is anti-Hermitian in
Euclidean space and hence has purely imaginary eigenvalues.
Note also that $\rho_A$ is implicitly a functional of the gauge configuration $A$
through the covariant derivative.

   A useful property of the Dirac operator is that it
anti-commutes with $\gamma_5$ which, in turn, implies that for every nonzero
eigenvector $q_j$, $\gamma_5 q_j$ is an eigenvector with opposite eigenvalue.
This, in turn, implies that $\rho_A(\lambda)$ is an even function of $\lambda$.
The parameter $\lambda$ will be referred to as the virtuality.
There is an extremely simple way to represent $\rho_A$ as a trace which we will find
useful later:
\begin{equation}
\rho_A (\lambda) \, = \, \lim_{\epsilon \rightarrow 0} \pi /{\rm Vol} \, \,
 {\rm Im} \left ( \, {\rm Tr} \left [\frac{1}{\lambda  + i D_{\mu} \gamma^{\mu} - i
 \epsilon} \right] \right)
\label{simple} 
\end{equation}
where  Tr indicates a  trace over Dirac indices and also a functional
trace over configurations.

The thermal expectation of $\rho_A$ will be
of prime interest. I will define $\rho(\lambda)$ to be expectation value:
\begin{equation}
\rho(\lambda) \, = \, \langle \rho(\lambda)_A \rangle
\end{equation}
It is not easy to gain an immediate intuitive grasp of the meaning of 
$\rho(\lambda)$.  However, as will be stressed in this talk, $\rho(\lambda)$ in 
the vicinity of $\lambda=0$ is intimately related to the spontaneous breaking
of chiral symmetry and anomalous breaking of $U(1)$ axial symmetry.

Before proceeding, let me remind you of some well-known facts about chiral
symmetry in QCD.  The up and down quark masses in QCD (which I will generically
call $m_q$), are much less than the typical
hadronic scales.  Neglecting the quark masses is a reasonable first
approximation which we will take here and the effects of nonzero quark masses
can be computed via chiral perturbation theory.  Now if we neglect quark masses
the QCD Lagrangian is invariant under
$$q \, \rightarrow  \, e^{i \gamma_5 \vec{\tau} \cdot \vec{\theta}/2}$$
where $\vec{\tau}$ are the Pauli matrices and $\vec{\theta}$ are arbitrary
parameters.  While the underlying theory has this symmetry, the ground state
({\it i.e.} the vacuum) does not, at least for $T<T_c$: 
the symmetry is spontaneously broken.  One easy way to see this is to note that
the chiral condensate $\langle \overline{q} {q} \rangle$ is nonzero.  This
indicates spontaneously  chiral symmetry breaking  since under a chiral
rotation through $\pi$, in the $a$ direction,
the operator $\overline{q} {q}$ transforms into 
$\overline{q} i \gamma_5 \tau_a q$ which has a zero expectation value.
The consequence of this spontaneous breaking are profound:  through Goldstone's
theorem  we see that the symmetry breaking implies the existence of massless
isovector pseudoscalars which we interpret as pions.  (In nature, they
are merely very light and not massless since the quark masses are not exactly 
zero).  Similarly, the spontaneous symmetry breaking implies that the pions don't
interact at threshold (in the zero quark mass limit) which allows one to
formulate a chiral perturbation theory for low energy 
interactions.~\cite{chipt}

It is generally believed that in the high temperature phase of QCD that chiral
symmetry is restored.  That is to say that the symmetry of the thermal state is
the same as the symmetry of the Lagrangian.  One consequence of this is above $T_c$
the chiral condensate vanishes.

Let me now turn to the connection between the chiral condensate and the
spectrum of the Dirac operator.  Banks and Casher proved a remarkable result in
1980:~\cite{BC}
\begin{equation}
\lim_{m_q \rightarrow 0} \, \, \langle \overline{q} q \rangle \, = \, \pi \,
\rho(0)
\label{BC}
\end{equation}
The chiral condensate in the zero quark mass limit is directly
proportional to the density of states at zero virtuality.
The proof of this  is quite straightforward:
\begin{equation}
\langle \overline{q} q(x=0)  \rangle \, = \,  \langle {\rm tr}[ S_A(0,0)] \rangle
\label{BC1}\end{equation}
where $S_A$ is the quark propagator in the presence of a background gluon
field,
\begin{equation}
(D_{\mu} \gamma^\mu + m_q) S_A(x,y) = \delta^4(x-y) \,,
\end{equation}
and tr indicates a trace over the Dirac indices only.
Since $S_A(x,y)$ is simply the inverse of the operator 
$D_{\mu} \gamma^\mu + m_q$, it is clear that spatially averaging over all 
$x$ will simply yield the functional trace of $(D_{\mu} \gamma^\mu +
m_q)^{-1}$:
$$\int \, d^4 x tr[S_A(x,x)] \, = \, Tr \left [\frac{1}{D_\mu \gamma^\mu +
m} \right]$$
One can exploit translational invariance to note the 
$$\langle \overline{q} q
\rangle \, = \,\langle \frac{1}{{\rm Vol}} \, \, \int \, d^4 x \, \overline{q} q (x)
\rangle \, = 
\, 1/{\rm Vol} \,  \langle Tr \left [\frac{1}{D_\mu \gamma^\mu +
m} \right]\rangle $$ 
Taking the limit as $m \rightarrow 0$ and comparing with eq.~(\ref{simple})
immediately yields the Banks-Casher relation of eq.~(\ref{BC}). 

One obvious and beautiful thing about the derivation is that it only depends
on the Euclidean functional integral formulation of the problem and hence is 
automatically valid at finite temperature. Above the phase transition
where chiral symmetry is restored and $\langle \overline{q} q \rangle$ goes to
zero (or more strictly ${\cal O}(m_q)$  ) we see that $\rho(0) \rightarrow 0
$ (or more precisely ${\cal O}(m_q)$).  So we see that $\rho(0)$ serves as an
order parameter for chiral symmetry and its restoration.  

The spectral density of the Dirac operator also gives us useful insights  into 
the  $U(1)_A$  problem ({\it i.e.} the axial $U(1)$) problem).  It is well known that
the QCD Lagrangian with massless quarks has an additional axial symmetry:
\begin{equation}
q \rightarrow e^{i \gamma_5 \theta/2} q
\end{equation} 
This symmetry is anomalously broken by quantum
effects---the current associated with it is not conserved.
The reason for this is quite simple: divergences in the theory require
regularization;  there is no way to regularize which simultaneously
respects $U(1)_A$ symmetry and gauge symmetry.  Since the renormalization of
the theory requires unbroken gauge symmetry, $U(1)_A$ must break.  The
consequences of this are far reaching.  For example, this is an essential 
ingredient in the understanding of why there is no $U(1)_A$ Goldstone boson.~\cite{c}

Now, just as there is a connection between the spectral density at zero
virtuality and the spontaneous breaking of ordinary chiral symmetry, there
is an important connection between the states at zero virtuality and the
spontaneous/anomalous breaking of the $U(1)_A$.  The point is 
very simple---all effects  associated with $U(1)_A$ breaking in massless QCD
can be traced to modes of  the Dirac operator 
in the neighborhood of zero virtuality.~\cite{cohen1}  To make this concrete,
consider what I will call as the $U(1)_A$ susceptibility, $\chi_{U(1)_A}$,
which is simply the spatial integral of the correlator in the isovector
pseudoscalar ($\pi$) channel minus correlator in the isoscalar pseudoscalar
(`` $ \eta^\prime $) '' channel:
\begin{equation} 
\chi_{U(1)_A} \, = \,  \int d^4 x \, \langle j_\pi(x) j_\pi(0) \, - \,
j_{\eta^\prime}(x) j_{\eta^\prime}(0) \rangle
\end{equation} 
with $j_\pi = \overline{q}\gamma_5 \tau q$ and $j_{\eta^\prime} = \overline{q}
\gamma_5 q$.  An analysis very similar to that of the Banks-Casher relation
immediately yields;
\begin{equation}
\chi_{U(1)_A} \, = \, 2 \int \, d \lambda \, \rho(\lambda) 
\frac{m_q^2}{\lambda^2 + m_q^2}
\label{chiU1}\end{equation}
>From the form of eq.~(\ref{chiU1}) it is obvious that in the limit $m_q
\rightarrow 0$ the susceptibility must vanish unless there is strength in the 
immediate neighborhood of zero.  The important thing to realize is that
as was pointed out a couple of years back~\cite{cohen1} this is generic: all
$U(1)_A$ violating amplitudes get their strength entirely from the
$\lambda \sim 0$ region.

A few years ago, motivated by an insight of the previous sort (in the context of the
instanton liquid model), Shuryak~\cite{s}
asked the following provocative question: since we know above $T_c$, $\rho(0)=0$ and we know that $U(1)_A$
violating amplitudes come from modes at $\lambda=0$, is it possible that all
effects of the anomalous  $U(1)_A$ breaking vanish in the chirally restored
phase?  At first sight this seems completely nuts---the anomaly is an operator equation and the axial  current is not
conserved regardless of the state of the system.  On the other hand, it is well
known that it is only due to subtle interplay between topology and the anomaly
that any effects of $U(1)_A$ breaking are seen. 
In fact, a couple of years ago
I  showed on very general grounds based on the positivity of the measure in the
QCD along the lines of the QCD inequalities of Weingarten, Witten and
Vafa~\cite{QCDIN} that
unless there are contributions to the functional integral which form a set of
measure zero in the massless limit of the theory, $U(1)_A$ violating amplitudes
must vanish above $T_c$.   This raises the obvious question as to whether such
a set of measure zero can contribute.  As noted by Lee and Hatsuda~\cite{LH} and
 by Evans,  Hsu and Schwetz~\cite{H},  such contributions are possible in a dilute instanton gas where
one might expect $\rho(\lambda) \sim m_u m_d \delta(\lambda)$.  Such a form
clearly is a set of measure zero in the massless quark limit but would contribute
to $U(1)_A$ violating amplitude.   This leaves us with the question of whether QCD
actually does yield $U(1)_A$ violating effects above $T_c$ or not.

Now the preceding analysis is highly formal and mathematical and  it is useful
to remind ourselves of what Bertrand Russell once said:
``{\it Mathematics may be defined as the subject in which we never know what
we are talking about.}''

Given the bleak prospects of using purely formal reasoning to answer this, it
seems there are two possible avenues of attack.   The first is through
phenomenology and the second is via lattice studies.

As I noted at the outset it is very difficult to use  phenomenology to pin
things down since so much of the dynamics is unknown.  There is at least one
scenario where we can get a dramatic effect but it depends on many unproven
assumptions.  The first is that $U(1)_A$ violating effects do vanish above
$T_c$; this is, of course, what we are trying to check.  The second is that the 
transition is second order and happens sufficiently rapidly to get out of
thermal equilibrium over large domains and develops an instability for pionic
growth.  This is the so-called disoriented chiral condensate scheme of Wilczek
and Rajagopal~\cite{dcc}.  If this happens, then along with collective
enhancements of low $P_T$ pions we would get enhancements of low $P_T$
$\eta^\prime$s.  Clearly this scenario is highly speculative and we turn to the
question of what the lattice can tell us.  

\section{Lattice Results}

In principle, the lattice  can answer our question about $U(1)_A$ violations
above $T_c$.  Of course, there may be the usual problems of interpretations
given effects of finite masses, lattice spacings and lattice sizes and the like.
 While there have been a number of lattice calculations of
quantities sensitive to $U(1)_A$ violations many of them depend on very
numerically unstable quantities such as screening masses in various channels
which we know are highly sensitive to quark mass effects and threshold effects.  A more natural
quantity to study is the $U(1)_A$ violating susceptibility defined in
eq.~(\ref{chiU1}).   To my knowledge there are only two calculations in the
literature of this quantity.  The first by Chandrasekharan and Christ~\cite{CC} and the second by Bernard {\it et al.}~\cite{B}.
Both of these papers claim that that they see evidence that $\chi_{U(1)_A}$
does not go to zero above $T_c$ indicating that $U(1)_A$ violating effects 
do not vanish above $T_c$. 
In particular, Chandrasekharan and Christ claim that they see a small but nonvanishing $\chi_{U(1)_A}$ for modestly small quark masses. 
Bernard {\it et al.}  did  a more systematic study; they calculate for a number of modestly small quark masses
and extrapolate back to zero and obtain a nonzero value.  In contrast, the ordinary $SU(2) \times SU(2)$
chiral symmetry breaking susceptibility (the pion--sigma channels) goes nicely
to zero.

To some this may seem to settle the question; but to quote Mark Twain, ``{\it It's
differences of opinion which make horse races.}''  There are several reasons 
to doubt the lattice studies.  The first is the obvious one; the calculation involves a small
number from an extrapolation of quantities with unknown and potentially large 
systematic errors due to finite masses, lattice spacings sizes {\it etc.}
Moreover at least one class of these systematic errors (finite quark masses
compounded by finite lattice spacing)
can easily account for a spurious $U(1)_A$ violating amplitude without
affecting ordinary chiral symmetry.  The reason is quite simple: due to the
fermion doubling problem, viable  lattice schemes in general violate axial symmetries in a
spurious way.  There are different schemes for dealing with fermion doubling
which all have the general feature that the spurious axial symmetry violating
effects vanish in the continuum limit of $a \rightarrow 0$.  The calculations by
Chandrasekharan and Christ and   Bernard {\it et al.} used staggered fermions.
These have the property that even at finite lattice spacing one axial symmetry
is conserved (for $m_q=0$).  This conserved axial current is the one which is
conventionally associated with the pion channel.  On the other hand, the
$U(1)_A$ symmetry is explicitly broken by the lattice formulation and is only
restored in the continuum limit.  Thus, for example, even  in the case of noninteracting
massless fermions in this lattice formulation one would see $U(1)_A$ violations
until $a$ is sent to zero.

Accordingly it would be very useful to study the anatomy of these calculations
to see whether the results  are coming from lattice artifacts.  As was
discussed in the introduction, $\chi_{U(1)_A}$ is completely determined by
$\rho(\lambda)$ near zero.  Thus if one could measure $\rho(\lambda)$ for the
lattice configurations contributing, one could see from where the $U(1)_A$
violating amplitudes get their strength.  In fact, as we will see, the $U(1)_A$
violations above $T_c$ in ref. 11
%\cite{CC} 
get their strength in a manner
which appears to be inconsistent
with chiral restoration suggesting the result may well be due to lattice artifacts.

While a direct lattice calculation of $\rho(\lambda)$ is what we want, there
are presently no available such calculations.  This is not surprising as it
would be a numerical nightmare to calculate.
Fortunately an integral transform
of $\rho(\lambda)$ has been calculated by Chandrasekharan and Christ.~\cite{CC}  In particular they calculate
\begin{equation}
f(m_\xi) \, = \, \int \, d \lambda \, \rho(\lambda) \, \frac{1}{i \lambda + m_\xi}
\label{a}\end{equation}
This is actually a relatively easy calculation to do on the lattice since it amounts to
calculating $\langle \overline{q} q \rangle$ with a quark mass of $m_\xi$ in
the propagator instead of $m_q$.  Hence standard codes can be used.  Since the quark mass
in the propagator can be made much less than that in the functional determinant
(which is implicitly contained in $\rho$) one can call such a calculation
partially quenched.  Note that by construction, $f(m_q) = \langle \overline{q} q
\rangle$.

Chandrasekharan and Christ calculated $f(m_\xi)$ 
for various choices of coupling constants which correspond to different values
of the temperature.  (Let me remind you that by changing the coupling constant
with fixed physical observables one is effectively changing the lattice
spacing.  Since the number of steps in the temporal direction is fixed, this in
turn changes the periodicity of the lattice in time and hence the temperature.)
The important thing for our purposes is that for temperatures above $T_c$ and small $m_\xi$, the function is  apparently linear on a log-log plot over several decades implying that
 $f(m_\xi) \sim m_\xi^\alpha$ for sufficiently small $\lambda$.
Moreover it is found that in all cases $\alpha <1$.

Now it is easy to see that this behavior implies
$$\rho(\lambda) = c |\lambda|^\alpha\,,$$
where $c$ is a constant, and this form
applies at sufficient small $\lambda$.
The simplest way to see this is to put this form into eq.~(\ref{a}) and
evaluate the integral. One finds that $\rho \sim |\lambda|^\alpha$ implying $f \sim
m_\xi^\alpha$.   Indeed this is seen easily by dimensional analysis.
In the following section, I will show that there is a constraint imposed by
chiral restoration which shows that a form $\rho \sim |\lambda|^\alpha$
with $\alpha < 1$ cannot occur.  This in turn raises questions as to whether
the lattice results are dominated by artifacts.

\section{Constraints}

Recall that we are studying the symmetry restored phase. In a chiral symmetric
phase one knows  that the log of the partition function  is an even analytic
function of the quark mass near zero.  The reason for this is quite
easy to understand.  Look at the $n^{\rm th}$ derivative of $\log (Z)$ with
respect to the quark masses:  
\begin{equation}
\left . \frac{\partial^n \log (Z)}{\partial m_q^n} \right |_{m_q=0}  \, = \, \frac{1}{V} \, \langle
\int \,( d^4 x \, \overline{q} q)^n \rangle \,.
\end{equation}
In the chiral restored phase this quantity should not change under a chiral
rotation. Under a chiral rotation through $\pi$, $\overline{q} q \, \rightarrow
\, - \overline{q} q$, thus the preceding expression must vanish for all odd n;
in general, for even $n$ we expect nonzero finite values implying that $\log
(Z)$ is an even analytic function of  $m_q$.
 
We should recall that the quark condensate is given by 
 \begin{equation}
 \langle \overline{q} q \rangle \, =
 \, \frac{1}{{\rm Vol}}\frac{\partial \log (Z)}{\partial m_q} \,,
\end{equation}
and that from the Banks-Casher analysis,
\begin{equation}
\langle \overline{q} q) \rangle \, = \int \, d \lambda \, \rho (\lambda)
\frac{m_q}{\lambda^2 + m_q^2} \,.
\end{equation}
Combining these we see that
\begin{equation}
\lim_{m_q \rightarrow 0}  \frac{1}{{\rm Vol}} \, \frac{\partial^n \log (Z)}{\partial m_q^n}   \,
= \, \frac{\partial^{n -1}}{\partial m_q^{n-1}} \,  \int \, d \lambda \, \rho
(\lambda)
\, \frac{m_q}{\lambda^2 + m_q^2}
\label{f1}\end{equation}
Recall that this must be  zero for all odd n.

Before proceeding with a careful analysis, let me begin with a small swindle.
For your convenience, I will  identify the swindle at the outset:
assume that in eq.~(\ref{f1}) the quark mass derivative acts only on 
${m_q}/{\lambda^2 + m_q^2}$ and not on $\rho(\lambda)$.  (I should remind
you that $\rho$ depends implicitly on $m_q$ through the functional
determinant.)  Diagrammatically, this  corresponds to including only
quark-line--connected diagrams.  I will label any quantity calculated at this
level with the superscript $qlc$.   From eq.~(\ref{f1}) we see that 
\begin{equation}
\lim_{m_q \rightarrow 0} \,   \left . \frac{\partial^n \log (Z)}{\partial m_q^n}
\right |^{\rm qlc}  \,
= \, \lim_{m_q \rightarrow 0} \,  \int \, d \lambda \, \rho (\lambda) \, 
\frac{\partial^{n - 1}}{\partial m_q^{n-1}} \, 
\frac{m_q}{\lambda^2 + m_q^2} \, = \, - \int \, d \, \lambda \, \rho (\lambda)
\frac{i}{\lambda^n} \,. \label{f2}
\end{equation}
Now if $\rho(\lambda) \sim |\lambda\| ^\alpha$ for small $\lambda$, then from
eq.~(\ref{f2}) one sees that the integral diverges for any $n > \alpha$.
On the other hand if $\log (Z)$ is an even analytic function this is impossible.
We conclude from this analysis the following: if only quark-line--connected
diagrams contribute, then $\rho$ (for $m_q=0$) is infinitely flat in $\lambda$---it goes to
zero faster than any power law.  This is what you would expect if $\rho$ had a
gap at zero.  By a gap at zero, I simply mean that  $\rho(\lambda) = 0$ unless
$\lambda$ is greater than some minimum.  Alternatively, one could imagine a
situation in which $\rho$ had an essential singularity at $\lambda=0$, such
as $\rho \sim e^{-1/\lambda}$.  In any case, $\rho$ is infinitely flat at the
origin.

My guess is that the preceding description is correct---that 
above $T_c$, $\rho$ does, in fact, have a gap.
I cannot  prove it however.  The problem is that the preceding
analysis only included the quark-line--connected diagrams.  It is possible in
principle  that the quark-line--disconnected parts, coming from derivatives of 
$m_q$ acting on $\rho(\lambda)$, conspire to cancel the divergent quark-line--connected
part.  One can prove, however, in the chiral restored phase,  that for $m_q=0$,  $\rho$ must go to zero
faster than linearly:
$\left .  \frac{\partial \rho}{\partial \lambda} \right |_{\lambda=0} =0$.

The proof of this constraint  is actually quite simple.  Consider the
susceptibility associated  with the pion channel,
\begin{equation}
\chi_{\pi} \, = \,  \int d^4 x \, \langle j_\pi(x) j_\pi(0)  \rangle \,,
\end{equation} 
with $j_\pi = \overline{q}\gamma_5 \tau q$.
By standard analysis of the propagator  analogous to the derivation of the
Banks-Casher result discussed in the introduction, one finds:
\begin{equation}
\chi_{\pi} \, = \,  \int \, d \lambda \frac{\rho(\lambda)}{\lambda^2 + m_q^2}
\, = \, \frac{\langle \overline{q}{q} \rangle}{m_q} 
\end{equation}
This last form is just a functional integral derivation of the Ward identity
implicit in the Gell-Mann--Oakes-Renner relation.
Now suppose that for small $\lambda$, $\rho = c | \lambda |^ \alpha$ with
$\alpha  \le 1$, and $c$ is a constant of proportionality.  It is a trivial exercise to  evaluate  
the integral in 
definition of $\chi_\pi$ using this form, and one finds that
\begin{equation}
 \chi_\pi \sim c m_q^{\alpha -1} . \label{chipi}
 \end{equation}
Let us suppose for the moment that the form $\rho = c |\lambda|^\alpha$
survives in the chiral  limit of $m_q \rightarrow 0$ so that $c$ is finite
in this limit.  Evaluating the integral in eq. (\ref{chipi}) for this case
gives for small $m_q$
\begin{equation}
\chi_\pi \, = \, c \, \frac{\pi}{m_q^{1-\alpha} \, \cos ( \alpha \pi/2)}  
\end{equation} 
which diverges in the chiral limit $m_q \rightarrow 0$,  provided $\alpha \le 1$. 
(For $\alpha =1$ it diverges as $\log (m_q)$.)
Moreover, if as
hypothesized, one were studying the chiral restored phase then as $m_q \rightarrow 0$
the susceptibilities in the pion and $\sigma$ channels must be equal.  On the
other hand 
\begin{equation}
\chi_\sigma \, = 
 \, \frac{1}{{\rm Vol}} \,  \frac{\partial^n \log (Z)}{\partial m_q^n} \, \, \,
 .
\end{equation}
Thus, if $\chi_\pi$ (and hence $\chi_{\sigma}$) diverge in the chiral limit
 we see that $\log(Z)$ is not an analytic function of $m_q$ in contradiction 
 to the system being in a chiral restored phase. 
 
>From the preceding analysis one can deduce that  in the chiral limit of
$m_q \rightarrow 0$ and the chirally restored phase, $\rho (\lambda)$ cannot
go as $c |\lambda|^\alpha$ for $\alpha \le 1$.  This appears to be inconsistent
with the lattice results of Chandrasekharan and Christ.  Of course, the
preceding analysis was done for the chiral limit of $m_q \rightarrow 0$, while
the lattice studies were done with  a fixed quark mass.  This raises the
following possibility. Suppose the constant $c$ used above were not a constant
but depended on the quark mass in such a way that it went to zero as $m_q
\rightarrow 0$. Would such a scenario allow a $U(1)_A$ violating susceptibility
while still being constant with the observation that $f(m_\xi) \sim
m_\xi^\alpha$?  At first sight it appears to be easy to do this:  
$f(\xi)\, \sim \, m_q^{1- \alpha} m_\xi^\alpha$, {\it i.e.} the constant $c$
could be proportional to $m_q$.  This possibility is consistent with the
lattice data in Chandrasekharan and Christ.  Since they do not do the
calculation for multiple values of $m_q$ at fixed $m_\xi$ it is impossible to
tell if this behavior is present.  If $f(m_\xi)$ behaves this way it is clear
 that for small $\lambda$ and $m_q$,
\begin{equation} 
\rho(\lambda) \, = \, b \,  m_q^{1 - \alpha}
|\lambda|^\alpha 
\label{rhod}
\end{equation}
with $b$ a constant.
If $\rho$ is given by eq.~(\ref{rhod}) it is a simple matter to verify that
$\chi_\pi$ and $\chi_\sigma$ are finite and equal and that $\chi_{U(1)_A}\, = \, 
(1 - \alpha) \chi_\pi$.  

The preceding scenario appears to show a way to reconcile the lattice
data with chiral restoration and $U(1)_A$ violation.  In fact, this scenario 
is inconsistent, unless there is spontaneous symmetry breaking of isospin and is
therefore not physically realized.  This  leaves an
inconsistency  which suggests that  the lattice results may be dominated by
unphysical lattice artifacts.  The issue of isospin has up until now
been ignored.  Implicitly I have taken the mass of the two light quarks to be
degenerate.  Assuming isospin is not spontaneously broken, this is not an issue
since at the end of the problem both the up and down quark masses are taken to
zero.  If isospin is spontaneously broken, the order in which the limit is
taken can matter, but if it is not spontaneously broken, then the ordering is irrelevant and we might
just as well take the masses equal.  The problem with the form in
eq.~(\ref{rhod}) is that it depends on the quantity $m_q^{1-\alpha}$.  If one 
generalizes to $m_u \ne m_d$ one must generalize  $m_q^{1-\alpha}$ into some
function of $m_u$ and $m_d$ which goes to $m_q^{1-\alpha}$ when $m_u=m_d =
m_q$.   Consider the following quantity:
\begin{eqnarray}
\chi_{u d} \, & =  & \, \int \, d^4 x \, \langle \overline{u}{u}(x) \, 
\overline{d}{d}(0) \rangle \nonumber \\
& = & \, \frac{1}{{\rm Vol}} \, 
\frac{\partial^2 \log (Z)}{\partial m_u \,  \partial m_d}
\end{eqnarray}
 Plugging in $\rho$ from eq.~(\ref{rhod})  into a generalization 
 of eq.~(\ref{f1}), for the case where $m_u$ and $m_d$ are varied
 independently, and evaluating the integral yields:
 \begin{equation}
\chi_{u d} \, \, = \,   
\frac{\partial m_q(m_u,m_d)^{1-\alpha}}{\partial m_d} \, \frac{b \pi \, m_u^\alpha}
{ \cos ( \alpha \pi/2)}  \, \, \, \, .
 \label{chiud2}\end{equation}
 
If isospin and chiral symmetry are both unbroken then the result for $\chi_{u
d}$ will be finite as $m_u \, , \,  m_d \, \rightarrow 0$ and is independent of how
the quark masses go to zero. Let us assume this to be true and show that it
leads to a contradiction.  On dimensional grounds we can always write
\begin{equation}
\frac{\partial m_q^{1-\alpha}}{\partial m_d} \,  =  \, (1- \alpha)
m_u^{-alpha} \, g(m_u/m_d)
\end{equation}
where $g$ is some presently unknown function.  Inserting this form into
eq.(\ref{chiud2}) gives
\begin{equation}
\chi_{u d}  \, = \,  g(m_u/m_d) \frac{b \pi \, m_u^\alpha}
{ \cos ( \alpha \pi/2)}  
\label{ud}\end{equation}
By hypothesis, isospin and chiral symmetry are not spontaneously broken so that
the value of
$\chi_{u d}$ must be the same 
independently of how the chiral limit is approached.  This in turn means that the
value of $\chi_{u,d}$ has to be independent of the ratio $m_u/m_d$ since we can
approach the chiral limit with this ratio fixed to any value we wish.   From
this we deduce that consistency requires $g=1$
and $m_q^{1 - \alpha} \, = \, m_d m_u^-\alpha$ which in turn implies
\begin{equation}
\rho(\lambda) \,  = \,  b \, m_d m_u^-\alpha \, |\lambda|^{\alpha}
\end{equation}
At this point, however, we see a contradiction:  the preceding expression
is not invariant under $m_u \leftrightarrow m_d$; on the other hand since theory
is isospin symmetric up to the values of the quark masses the spectral density
must be invariant under the switching of the up and down quark masses.
Thus we conclude that the scenario in which $\rho(\lambda) \sim m_q^{1- \alpha}
|\lambda|^{\alpha}$ is not consistent.

\section{Summary}

Let  me summarize the situation:  I have discussed the concept of the spectral
density near $\lambda =0$, and argued that it was interesting and an important theoretical tool for
understanding both spontaneous chiral symmetry breaking and the
spontaneous/anomolous breaking of $U(1)_A$.   One problem of physical interest
that this tool may prove useful for is the question of whether the effects
of $U(1)_A$. axial symmetry breaking are manifest above the chiral phase
transition.  Although present lattice simulations seem to indicate that
$U(1)_A$ violating effects survive above the transition, it is possible
that the calculations are dominated by lattice artifacts.  One way to see
if this is so, is to test whether the spectral densities implied by the
lattice calculation are consistent with constraints imposed on the spectral
density by chiral symmetry and isospin in the unbroken phase.

There are a number of such  constraints: first, one sees that unless
quark-line--disconnected graphs conspire to cancel the quark-line--connected 
ones, in the chiral limit, $\rho(\lambda)$ is infinitely flat in the sense
of having all derivatives vanish as one would expect from a spectrum with a
gap.  Second, in the chiral limit, 
$\frac{\partial \rho}{\partial \lambda}=0$,
indicating  a behavior such as $\rho(\lambda, m_q =0) \sim 
|\lambda|^{\alpha}$,
is not possible for $\alpha \le 1$.  Finally it is shown that for small $\lambda$ and $m_q$
a behavior of the form
$\rho(\lambda, m_q =0) \sim m_q^{1-\alpha} |\lambda|^{\alpha}$ is not possible
for $\alpha \le 1$.

Given these constraints it is reasonable  to ask whether the lattice
calculation of  Chandrasekharan and Christ, which have $\rho$ behaving like a power law with
$\alpha <1$ (and hence in apparent violation of the constraint) are dominated 
by lattice artifacts.   Since the lattice calculations of ref. 12
%\cite{B} 
do not
study the spectral density near zero it is impossible to know whether they are
consistent with the constraints derived above.  However, as these calculations
were qualitatively similar to those in ref. 11
%\cite{CC} 
it  is reasonable to
question whether the spectral functions are consistent with being in a chirally
restored phase.
I conclude with a
word of warning to potential consumers of lattice calculations for these
problems: Let the buyer 
beware.

\section*{Acknowledgements}

The financial assistance of the U.S. Department of Energy is greatly
acknowledged as is the hospitality of APCTP.

\end{document}